\documentclass[letterpaper, 10 pt, conference]{ieeeconf}%
\usepackage{makeidx}
\usepackage{graphicx}
\usepackage{multicol}
\usepackage[bottom]{footmisc}
\usepackage{latexsym}
\usepackage{subfigure}
\usepackage{cite,color,comment,xspace}
\usepackage{amsfonts}
\usepackage{amsmath}
\usepackage{amssymb}
\usepackage{algorithm}
\usepackage{algorithmic}
\usepackage{url}
\usepackage{times}
\usepackage{mathptmx}
\usepackage{enumerate}
\usepackage{epstopdf}
\usepackage{epsfig}%
\IEEEoverridecommandlockouts
\newtheorem{proposition}{\bf Proposition}

\newtheorem{remark}{\bf Remark}
\makeindex
\begin{document}

\title{An Optimal Control Approach for the Persistent Monitoring Problem}
\author{Christos G. Cassandras, Xu Chu Ding, Xuchao Lin \thanks{This work was
supported in part by NSF under Grant EFRI-0735974, by AFOSR under grants
FA9550-07-1-0361 and FA9550-09-1-0095, by DOE under grant DE-FG52-06NA27490,
and by ONR under grant N00014-09-1-1051.} \thanks{The authors are with the
Division of Systems Engineering and Center for Information and Systems
Engineering, Boston University, Boston, MA;
e-mail:\emph{\{cgc,xcding,mmxclin\}@bu.edu}.} }
\maketitle

\begin{abstract}
We propose an optimal control framework for persistent monitoring problems
where the objective is to control the movement of mobile agents to minimize an
uncertainty metric in a given mission space. For a single agent in a
one-dimensional space, we show that the optimal solution is obtained in terms
of a sequence of switching locations, thus reducing it to a parametric
optimization problem. Using Infinitesimal Perturbation Analysis (IPA) we
obtain a complete solution through a gradient-based algorithm. We also discuss
a receding horizon controller which is capable of obtaining a near-optimal
solution on-the-fly.
We illustrate our approach with numerical examples.

\end{abstract}

\thispagestyle{empty} \pagestyle{empty}

\section{Introduction}

Enabled by recent technological advances, the deployment of autonomous agents
that can cooperatively perform complex tasks is rapidly becoming a reality. In
particular, there has been considerable progress reported in the literature on
sensor networks that can carry out coverage control
\cite{rekleitis2004limited,cortes2004coverage,li2006cooperative}, surveillance
\cite{girard2005border,grocholsky2006cooperative} and environmental sampling
\cite{RNS-MS-SLS-DR-GSS:10f-JFR,paley2008cooperative} missions. In this paper, we are
interested in generating optimal control strategies for \emph{persistent
monitoring} tasks; these arise when agents must monitor a dynamically changing
environment which cannot be fully covered by a stationary team of available
agents. Persistent monitoring differs from traditional coverage tasks due to
the perpetual need to cover a changing environment, i.e., all areas of the
mission space must be visited infinitely often. The main challenge in
designing control strategies in this case is in balancing the presence of
agents in the changing environment so that it is optimally covered over time
while still satisfying sensing and motion constraints. Examples of persistent
monitoring missions include surveillance in a museum to prevent unexpected
events or thefts, unmanned vehicles for border patrol missions, and
environmental applications where routine sampling of an area is involved.

In this paper, we address the persistent monitoring problem through an
optimal control framework to drive agents so as to minimize a metric of
uncertainty over the environment. In coverage control
\cite{cortes2004coverage,li2006cooperative}, it is common to model knowledge
of the environment as a non-negative density function defined over the mission
space, and usually assumed to be fixed over time. However, since persistent
monitoring tasks involve dynamically changing environments, it is natural to
extend it to a function of both space and time to model uncertainty in the
environment. We assume that uncertainty at a point grows in time if it is not
covered by any agent sensors; for simplicity, we assume this growth is linear.
To model sensor coverage, we define a probability of detecting events at each
point of the mission space by agent sensors. Thus, the uncertainty of the
environment decreases (for simplicity, linearly) with a rate proportional to
the event detection probability, i.e., the higher the sensing
effectiveness is, the faster the uncertainty is reduced..

While it is desirable to track the value of uncertainty over all points in the
environment, this is generally infeasible due to computational complexity and
memory constraints. Motivated by polling models in queueing theory, e.g.,
spatial queueing \cite{bertsimas1993stochastic},\cite{cooper1981introduction},
and by stochastic flow models \cite{sun2004perturbation}, we assign sampling
points of the environment to be monitored persistently (equivalently, we
partition the environment into a discrete set of regions). We associate to
these points \textquotedblleft uncertainty queues\textquotedblright\ which are
visited by one or more servers. The growth in uncertainty at a sampling point
can then be viewed as a flow into a queue, and the reduction in uncertainty
(when covered by an agent) can be viewed as the queue being visited by mobile
servers as in a polling system. Moreover, the service flow rates depend on the
distance of the sampling point to nearby agents. From this point of view, we
aim to control the movement of the servers (agents) so that the total
accumulated \textquotedblleft uncertainty queue\textquotedblright\ content is minimized.

Control and motion planning for agents performing persistent monitoring tasks
have been studied in the literature. In \cite{rekleitis2004limited} the focus
is on sweep coverage problems, where agents are controlled to sweep an area.
In \cite{SLS-MS-DR:10a,nigam2008persistent} a similar metric of
uncertainty is used to model knowledge of a dynamic environment. In
\cite{nigam2008persistent}, the sampling points in a 1-D environment are
denoted as cells, and the optimal control policy for a two-cell problem is
given. Problems with more than two cells are addressed by a heuristic policy.
In \cite{SLS-MS-DR:10a}, the authors proposed a stabilizing speed
controller for a single agent so that the accumulated uncertainty over a set
of points along a given path in the environment is bounded, and an optimal
controller that minimizes the maximum steady-state uncertainty over points of
interest, assuming that the agent travels along a closed path and does not
change direction. The persistent monitoring problem is also related to robot
patrol problems, where a team of robots are required to visit points in the
workspace with frequency constraints
\cite{hokayem2008persistent,elmaliach2008realistic,elmaliach2007multi}.

Our ultimate goal is to optimally control a team of cooperating agents in a 2
or 3-D environment. The contribution of this paper is to take a first step
toward this goal by formulating and solving an optimal control problem for one
agent moving in a 1-D mission space in which we minimize the accumulated
uncertainty over a given time horizon and over an arbitrary number of sampling
points. Even in this simple case, determining a complete explicit solution is
computationally hard. However, we show that the optimal trajectory of the
agent is to oscillate in the mission space: move at full speed, then switch
direction before reaching either end point. Thus, we show that the solution is
reduced to a parametric optimization problem over the switching points for
such a trajectory. We then use generalized Infinitesimal Perturbation Analysis
(IPA) \cite{cassandras2009perturbation},\cite{Wardietal09}\textbf{ }to
determine these optimal switching locations, which fully characterize the
optimal control for the agent. This establishes the basis for extending this
approach, first to multiple agents and then to a 2-dimensional mission
space.\ It also provides insights that motivate the use of a receding horizon
approach for bypassing the computational complexity limiting real-time control
actions. These next steps are the subject of ongoing research.

The rest of the paper is organized as follows. Section
\ref{sec:problemformulation} formulates the optimal control problem. Section
\ref{sec:optimalsolution} characterizes the solution of the optimal control
problem in terms of switching points in the mission space, and includes IPA in
conjunction with a gradient-based algorithm to compute the sequence of optimal
switching locations. Section \ref{sec:exper} provides some numerical results.
Section \ref{sec:extensions} discusses extensions of this result to a receding
horizon framework and to multiple agents. Section \ref{sec:concl} concludes
the paper.

\section{Persistent Monitoring Problem Formulation}

\label{sec:problemformulation} We consider a mobile agent in a 1-dimensional
mission space of length $L$. Let the position of the agent be $s(t)\in\left[
0,L\right]  $ with dynamics:
\begin{equation}
\dot{s}(t)=u(t),\text{ \ \ }s(0)=0 \label{eq:agentdyn}%
\end{equation}
i.e., we assume that the agent can control its direction and speed.
We assume that the speed is constrained by $\left\vert u\left(
t\right)  \right\vert \leq1$.

We associate with every point $x\in\left[  0,L\right]  $ a function $p(x,s)$
at state $s(t)$ that captures the probability of detecting an event at this
point. We assume that $p(x,s)=1$ if $x=s$, and that $p(x,s)$ decays when the
distance between $x$ and $s$ (i.e.,\textit{ } $|x-s|$) increases. Assuming a
finite sensing range $r$, we set $p(x,s)=0$ when $|x-s|>r$.
In this paper, we use a linear decay model shown below as our event detection
probability function:
\begin{equation}
p(x,s)=\left\{
\begin{array}
[c]{ll}%
1-\frac{\left\vert x-s\right\vert }{r} & \text{if }|x-s|\text{ }\leq r\\
0 & \text{if }|x-s|\text{ }>r
\end{array}
\right.  \label{eq:linearmodel}%
\end{equation}

We consider a set of points $\{\alpha_{i}\}$, $i=1,\ldots,M$, $\alpha_{i}%
\in\lbrack0,L]$, and associate a time-varying measure of uncertainty with each
point $\alpha_{i}$, which we denote by $R_{i}(t)$. Without loss of generality,
we assume $0\leq\alpha_{1}\leq\cdots\leq\alpha_{M}\leq L$ and, to simplify
notation, we set $p_{i}(s(t))\equiv p(\alpha_{i},s(t)).$ This set may be
selected to contain points of interest in the environment, or sampled points
from the mission space. Alternatively, we may consider a partition of $[0,L]$
into $M$ intervals whose center points are $\alpha_{i}=(2i-1)L/2M$,
$i=1,\ldots,M$. \ We can then set $p(x,s)=p_{i}(s)$ for all $x\in\lbrack
\alpha_{i}-\frac{L}{2M},\alpha_{i}+\frac{L}{2M}]$. The uncertainty functions
$R_{i}(t)$ are defined to have the following properties: $(i)$ $R_{i}(t)$
increases with a fixed rate dependent on $\alpha_{i}$, if $p_{i}(s(t))=0$, $(ii)$ $R_{i}(t)$ decreases
with a fixed rate if $p_{i}(s(t))=1$, and $(iii)$ $R_{i}(t)\geq0$ for all $t$.
It is then natural to model uncertainty so that its decrease is proportional
to the probability of detection. In particular, we model the dynamics of
$R_{i}(t)$, $i=1,\ldots,M$, as follows:
\begin{equation}
\dot{R}_{i}(t)=\left\{
\begin{array}
[c]{ll}%
0 & \text{if }R_{i}(t)=0,\text{ }A_{i}<Bp_{i}(s(t))\\
A_{i}-Bp_{i}(s(t)) & \text{otherwise}%
\end{array}
\right.  \label{eq:dynamicsR}%
\end{equation}
where we assume that initial conditions $R_{i}(0)$, $i=1,\ldots,M$, are given
and that $B>A_{i}>0$ for all $i$ (thus, the uncertainty strictly decreases when
$s(t)=\alpha_{i}$).

\begin{figure}[th]
\center
\includegraphics[scale=0.45]{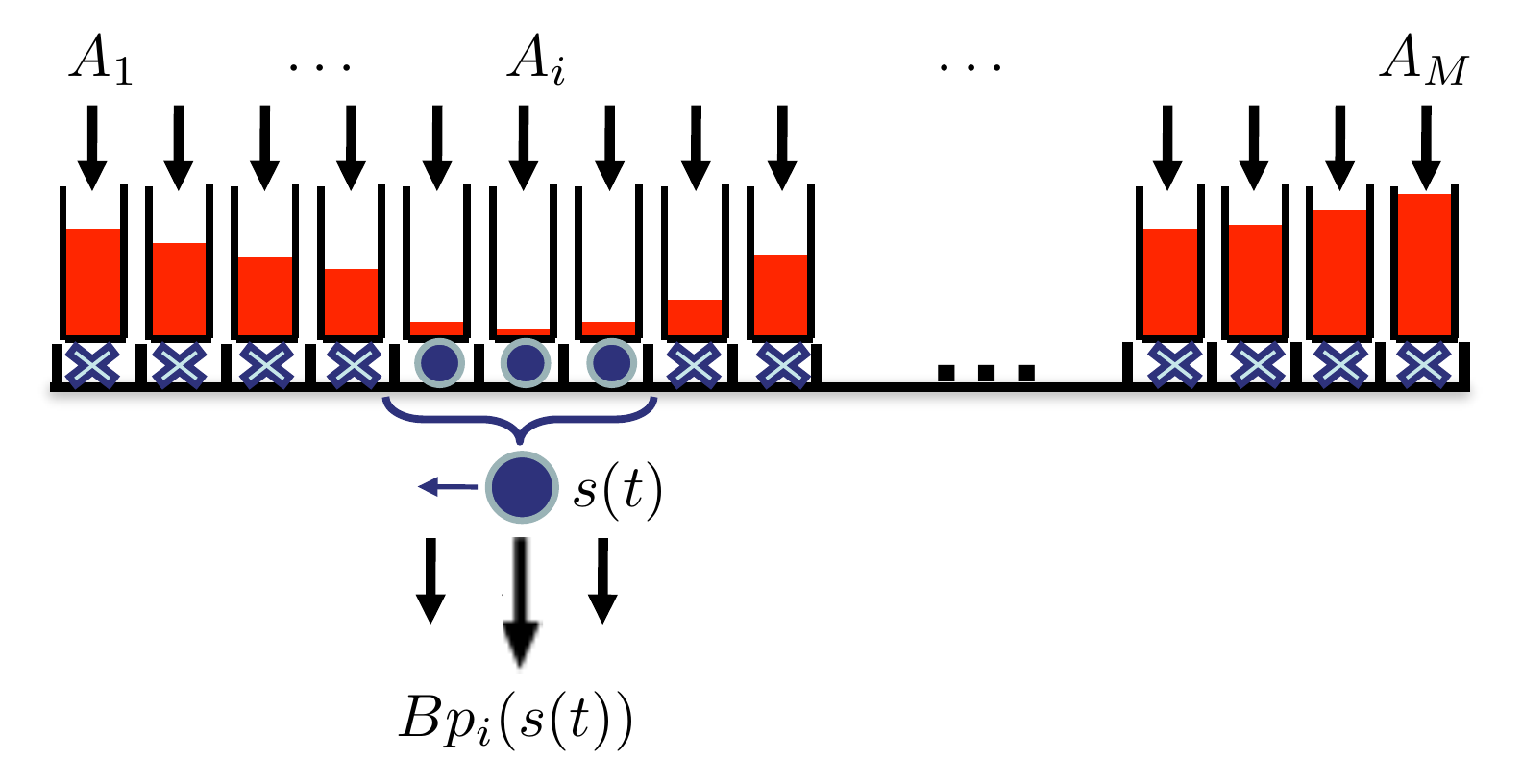}\caption{A queueing system analog of
the persistent monitoring problem.}%
\label{fig:queue}%
\end{figure}

Viewing persistent monitoring as a polling system, each point $\alpha_{i}$
(equivalently, $i$th interval in $[0,L]$) is associated with a
\textquotedblleft virtual queue\textquotedblright\ where uncertainty
accumulates with inflow rate $A_{i}$. The service rate of this queue is
time-varying and given by $Bp_{i}(s(t))$, controllable through the agent
position at time $t$, as shown in Fig. \ref{fig:queue}.  This interpretation is
convenient for characterizing the \emph{stability} of this system: For each
queue, we may require that $A_{i}<\frac{1}{T}\int_{0}^{T}Bp_{i}(s(t))dt$.
Alternatively, we may require that each queue becomes empty at least once over
$[0,T]$. We may also impose conditions such as $R_{i}(T)\leq R_{\max}$ for
each queue as additional constraints for our problem so as to provide bounded
uncertainty guarantees, although we will not do so in this paper. Note that
this analogy readily extends to multi-agent and 2 or 3-D settings.  Also, note that $B$ can also be
made location dependent without affect the analysis in this paper.

The goal of the optimal persistent monitoring problem we consider is to
control the mobile agent direction and speed $u(t)$ so that the cumulative
uncertainty over all sensor points $\{\alpha_{i}\},$ $i=1,\ldots,M$ is
minimized over a fixed time horizon $T$. Thus, we aim to solve the following
optimal control problem:
\begin{equation}
\text{\textbf{Problem P1}: \ \ \ \ }\min_{u\left(  t\right)  }\text{ }%
J=\frac{1}{T}\int_{0}^{T}\sum_{i=1}^{M}R_{i}(t)dt \label{eq:costfunction}%
\end{equation}
subject to the agent dynamics \eqref{eq:agentdyn}, uncertainty dynamics
\eqref{eq:dynamicsR}, state constraint $0\leq s(t)\leq L$, $t\in\lbrack0,T]$,
and control constraint $|u(t)|\leq1$, $t\in\lbrack0,T]$.

\section{Optimal Control Solution}

\label{sec:optimalsolution} In this section we first characterize the optimal
control solution of \textbf{Problem P1 }and show that it is reduced to a
parametric optimization problem. This allows us to utilize the IPA method
\cite{cassandras2009perturbation} to find a complete optimal solution.

\subsection{Hamiltonian analysis}

\label{sec:sec:hamiltonian} We define the state vector $x\left(  t\right)
=[s\left(  t\right)  ,R_{1}(t),\ldots,R_{M}(t)]^{\mathtt{T}}$ and the
associated costate vector $\lambda\left(  t\right)  =[\lambda_{s}\left(
t\right)  ,\lambda_{1}(t),\ldots,\lambda_{M}(t)]^{\mathtt{T}}$. In view of the
discontinuity in the dynamics of $R_{i}(t)$ in (\ref{eq:dynamicsR}), the
optimal state trajectory may contain a boundary arc when $R_{i}(t)=0$ for any
$i$; otherwise, the state evolves in an interior arc. We first analyze the
system operating in such an interior arc. Due to \eqref{eq:agentdyn} and
(\ref{eq:dynamicsR}), the Hamiltonian is:%
\begin{equation}
H\left(  x,\lambda,u\right)  =\sum_{i=1}^{M}R_{i}\left(  t\right)
+\lambda_{s}\left(  t\right)  u\left(  t\right)  +\sum_{i=1}^{M}\lambda
_{i}\left(  t\right)  (A_{i}-Bp_{i}(s)) \label{Hamiltonian}%
\end{equation}
and the costate equations $\dot{\lambda}=-\frac{\partial H}{\partial x}$ are:%
\begin{align}
\dot{\lambda}_{s}\left(  t\right)   &  =-\frac{\partial H}{\partial s}%
=-B\sum_{i=1}^{M}\lambda_{i}\left(  t\right)  \frac{\partial p_{i}%
(s)}{\partial s}\nonumber\\
&  =-\frac{B}{r}\sum_{i\in F^{-}(t)}\lambda_{i}(t)+\frac{B}{r}%
\sum_{i\in F^{+}(t)}\lambda_{i}(t)\nonumber\\
\dot{\lambda}_{i}\left(  t\right)   &  =-\frac{\partial H}{\partial R_{i}%
}=-1\text{, \ \ \ \ }i=1,\ldots,M, \label{dyn of costate s}%
\end{align}
where we have used (\ref{eq:linearmodel}), and the sets $F^{-}(t)$ and
$F^{+}(t)$ are defined as:%
\[
\left\{
\begin{array}
[c]{l}%
F^{-}(t)=\{i:s\left(  t\right)  -r\leq\alpha_{i}\leq s\left(  t\right)  \}\\
F^{+}(t)=\{i:s\left(  t\right)  <\alpha_{i}\leq s\left(  t\right)  +r\},
\end{array}
\right.
\]
so that they identify all points $\alpha_{i}$ within the agent's sensing
range. Since we impose no terminal state constraints, the boundary conditions
are $\lambda_{s}\left(  T\right)  =0$ and $\lambda_{i}\left(  T\right)  =0$,
$i=1,\ldots,M$. Applying the Pontryagin minimum principle to
(\ref{Hamiltonian}) with $u^{\star}(t)$, $t\in\lbrack0,T)$, denoting an
optimal control, we have%
\[
H\left(  x^{\star},\lambda^{\star},u^{\star}\right)  =\min_{u\in\lbrack
-1,1]}H\left(  x,\lambda,u\right)
\]
and it is immediately obvious that it is necessary for an optimal control to
satisfy:
\begin{equation}
u^{\star}(t)=\left\{
\begin{array}
[c]{ll}%
1 & \text{ if }\lambda_{s}\left(  t\right)  <0\\
-1 & \text{ if }\lambda_{s}\left(  t\right)  >0
\end{array}
\right.  \label{u*}%
\end{equation}
This condition excludes the case where $\lambda_{s}\left(  t\right)  =0$ over
some finite \textquotedblleft singular intervals\textquotedblright%
\ \cite{bryson1975applied}. It turns out this can arise only in some
pathological cases which we shall not discuss in this paper.

The implication of (\ref{dyn of costate s}) with $\lambda_{i}\left(  T\right)
=0$ is that $\lambda_{i}\left(  t\right)  =T-t$ for all $t\in\lbrack0,T]$ and
all $i=1,\ldots,M$ and $\lambda_{i}\left(  t\right)  $ is monotonically
decreasing starting with $\lambda_{i}\left(  0\right)  =T$. However, this is
only true if the entire optimal trajectory is an interior arc, i.e., the state
constraints remain inactive. On the other hand, looking at
(\ref{dyn of costate s}), observe that when the two end points, $0$ and $L$,
are not within the range of the agent, we have $\left\vert F^{-}(t)\right\vert
=\left\vert F^{+}(t)\right\vert $, since the number of indices $i$ satisfying
$s\left(  t\right)  -r\leq\alpha_{i}\leq s\left(  t\right)  $ is the same as
that satisfying $s\left(  t\right)  <\alpha_{i}\leq s\left(  t\right)  +r$.
Consequently, $\dot{\lambda}_{s}\left(  t\right)  =0$, i.e., $\lambda
_{s}\left(  t\right)  $ remains constant as long as this condition is
satisfied and, in addition, none of the state constraints $R_{i}(t)\geq0$,
$i=1,\ldots,M$, is active.

Thus, as long as the optimal trajectory is an interior arc and $\lambda
_{s}\left(  t\right)  <0$, the agent moves at maximal speed $u^{\star}\left(
t\right)  =1$ in the positive direction towards the point $s=L$. If
$\lambda_{s}\left(  t\right)  $ switches sign before any of the state
constraints $R_{i}(t)\geq0$, $i=1,\ldots,M$, becomes active or the agent
reaches the end point $s=L$, then $u^{\star}\left(  t\right)  =-1$ and the
agent reverses its direction or, possibly, comes to rest. In what follows, we
examine the effect of the state constraints and will establish the fact that
the complete solution of this problem boils down to determining a set of
switching locations over $(0,L)$ with the end points being infeasible on an
optimal trajectory.

The dynamics in (\ref{eq:dynamicsR}) indicate a discontinuity arising when the
condition $R_{i}(t)=0$ is satisfied while $\dot{R}_{i}(t)=A_{i}-B%
p_{i}(s(t))<0$ for some $i=1,\ldots,M$. Thus, $R_{i}=0$ defines an interior
boundary condition which is not an explicit function of time. Following
standard optimal control analysis \cite{bryson1975applied}, if this condition
is satisfied at time $t$ for some $k\in\{1,\ldots,M\}$,
\begin{equation}
H\left(  x(t^{-}),\lambda(t^{-}),u(t^{-})\right)  =H\left(  x(t^{+}%
),\lambda(t^{+}),u(t^{+})\right)  \label{eq:hamcontinu}%
\end{equation}
where we note that one can make a choice of setting the Hamiltonian to be
continuous at the entry point of a boundary arc or at the exit point. Using
(\ref{Hamiltonian}) and (\ref{eq:dynamicsR}), \eqref{eq:hamcontinu} implies:%
\begin{equation}
\lambda_{s}^{\star}\left(  t^{-}\right)  u^{\star}\left(  t^{-}\right)
+\lambda_{k}^{\star}\left(  t^{-}\right)  \left(  A_{i}-Bp_{k}%
(s(t^{-}))\right)  =\lambda_{s}^{\star}\left(  t^{+}\right)  u^{\star}\left(
t^{+}\right)  \label{Ham jump R equal to zero_2}%
\end{equation}
In addition, $\lambda_{s}^{\star}\left(  t^{-}\right)  =\lambda_{s}^{\star
}\left(  t^{+}\right)  $ and $\lambda_{i}^{\star}\left(  t^{-}\right)
=\lambda_{i}^{\star}\left(  t^{+}\right)  $ for all $i\neq k$, but
$\lambda_{k}^{\star}\left(  t\right)  $ may experience a discontinuity so
that:%
\begin{equation}
\lambda_{k}^{\star}\left(  t^{-}\right)  =\lambda_{k}^{\star}\left(
t^{+}\right)  -\pi_{k} \label{costate discont}%
\end{equation}
where $\pi_{k}\geq0$. Recalling (\ref{u*}), since $\lambda_{s}^{\star}\left(
t\right)  $ remains unaffected, so does the optimal control, i.e., $u^{\star
}(t^{-})=u^{\star}(t^{+})$. Moreover, since this is an entry point of a
boundary arc, it follows from (\ref{eq:dynamicsR}) that $\dot{R}_{k}%
(t^{-})=A_{i}-Bp_{k}(s(t^{-}))<0$. Therefore,
(\ref{Ham jump R equal to zero_2}) and (\ref{costate discont}) imply that
\[
\lambda_{k}^{\star}\left(  t^{-}\right)  =0,\text{ \ }\lambda_{k}^{\star
}\left(  t^{+}\right)  =\pi_{k}\geq0.
\]

The actual evaluation of the costate vector over the interval $[0,T]$ requires
solving (\ref{dyn of costate s}), which in turn involves the determination of
all points where the state variables $R_{i}(t)$ reach their minimum feasible
values $R_{i}(t)=0$, $i=1,\ldots,M$. This generally involves the solution of a
two-point-boundary-value problem. However, our analysis thus far has already
established the structure of the optimal control (\ref{u*}) which we have seen
to remain unaffected by the presence of boundary arcs where $R_{i}(t)=0$ for
one or more $i=1,\ldots,M$. Let us now turn our attention to the constraints
$s(t)\geq0$ and $s(t)\leq L$. The following proposition asserts that neither
of these can become active on an optimal trajectory.

\begin{proposition}
\label{lem:switchingpoints} On an optimal trajectory, $s^{\star}\left(
t\right)  \neq0$ and $s^{\star}\left(  t\right)  \neq L$ for all $t\in\left(
0,T\right]  .$
\end{proposition}

\begin{proof}
Suppose $s(t)\geq0$ becomes active at some $t\in(0,T)$. In this
case, $\lambda_{i}\left(  t^{-}\right)  =\lambda_{_{i}}\left(  t^{+}\right)  $
for all\ $i=1,\ldots,M$, but $\lambda_{s}\left(  t\right)  $ may experience a
discontinuity so that%
\[
\lambda_{s}\left(  t^{-}\right)  =\lambda_{s}\left(  t^{+}\right)  -\pi_{0}%
\]
where $\pi_{0}\geq0$ is a scalar constant. Since the constraint $s=0$ is not
an explicit function of time, \eqref{eq:hamcontinu} holds and, using (\ref{Hamiltonian}), we get%
\begin{equation}
\lambda_{s}^{\star}\left(  t^{-}\right)  u^{\star}\left(  t^{-}\right)
=\lambda_{s}^{\star}\left(  t^{+}\right)  u^{\star}\left(  t^{+}\right)
\label{derived costate equation}%
\end{equation}
Clearly, as the agent approaches $s=0$ at time $t$, we must have $\dot
{s}^{\star}(t^{-})=u^{\star}(t^{-})<0$ and, from (\ref{u*}), $\lambda_{s}^{\star
}\left(  t^{-}\right)  >0$. It follows that $\lambda_{s}^{\star}\left(
t^{+}\right)  =\lambda_{s}^{\star}\left(  t^{-}\right)  +\pi_{0}>0$. On the
other hand, $u^{\star}\left(  t^{+}\right)  \geq0$, since the agent must either
come to rest or reverse its motion at $s=0$, hence $\lambda_{s}^{\star}\left(
t^{+}\right)  u^{\star}\left(  t^{+}\right)  \geq0$. From
(\ref{derived costate equation}), this contradicts the fact that $\lambda
_{s}^{\star}\left(  t^{-}\right)  u^{\star}\left(  t^{-}\right)  <0$ and we
conclude that $s^{\star}(t)=0$ can not occur. By the exact same argument, $s^{\star}\left(  t\right)  =L$ also cannot occur.
\end{proof}

Based on this analysis, the optimal control in (\ref{u*}) depends entirely on
the points where $\lambda_{s}\left(  t\right)  $ switches sign and, in light
of Prop. \ref{lem:switchingpoints}, the solution of the problem reduces to the
determination of a parameter vector $\theta=[\theta_{1},\ldots,\theta
_{N}]^{\mathtt{T}}$, where $\theta_{j}\in(0,L)$ denotes the $j$th location
where the optimal control changes sign. Note that $N$ is generally not known a
priori and depends on the time horizon $T$.

Since $s(0)=0$, from Prop. \ref{lem:switchingpoints} we have $u^{\star}(0)=1$,
thus $\theta_{1}$ corresponds to the optimal control switching from $1$ to
$-1$. Furthermore, $\theta_{j},j$ odd, always correspond to $u^{\star}(t)$
switching from $1$ to $-1$, and vice versa if $j$ is even. Thus, we have the
following constraints on the switching locations for all $j=2,\ldots,N$:%
\begin{equation}
\left\{
\begin{array}
[c]{l}%
\theta_{j}\leq\theta_{j-1},\text{ if }j\text{ is even}\\
\theta_{j}\geq\theta_{j-1},\text{ if }j\text{ is odd}.
\end{array}
\right.  \label{eq:thetaconstraint}%
\end{equation}

It is now clear that the behavior of the agent under the optimal control
policy (\ref{u*}) is that of a hybrid system whose dynamics undergo switches
when $u^{\star}\left(  t\right)  $ changes between $1$ and $-1$ or when
$R_{i}(t)$ reaches or leaves the boundary value $R_{i}=0$. As a result, we are
faced with a parametric optimization problem for a system with hybrid
dynamics. This is a setting where one can apply the generalized theory of
Infinitesimal Perturbation Analysis (IPA) in \cite{cassandras2009perturbation}%
,\cite{Wardietal09} to obtain the gradient of the objective function $J$ in
(\ref{eq:costfunction}) with respect to the vector $\theta$ and, therefore,
determine an optimal vector $\theta^{\star}$ through a gradient-based
optimization approach.

\begin{remark}
If the agent dynamics are replaced by a model such as $\dot{s}(t)=g(s)+bu(t)$,
observe that (\ref{u*}) still holds, as does Prop. \ref{lem:switchingpoints}.
The only difference lies in (\ref{dyn of costate s}) which would involve a
dependence on $\frac{dg(s)}{ds}$ and further complicate the associated
two-point-boundary-value problem. However, since the optimal solution is also
defined by a parameter vector $\theta=[\theta_{1},\ldots,\theta_{N}%
]^{\mathtt{T}}$, we can still apply the IPA approach presented in the next section.
\end{remark}

\subsection{Infinitesimal Perturbation Analysis (IPA)}

Our analysis has shown that, for an optimal trajectory, the agent always moves
at full speed and never reaches either boundary point, i.e., $0<s^{\star
}(t)<L$ (excluding certain pathological cases as mentioned earlier.) Thus, the
agent's movement can be parametrized through $\theta=[\theta_{1}%
,\ldots,\theta_{N}]^{\mathtt{T}}$ where $\theta_{i}$ is the $i$th control
switching point and the solution of Problem \textbf{P1} reduces to the
determination of an optimal parameter vector $\theta^{\star}$. As we pointed
out, the agent's behavior on an optimal trajectory defines a hybrid system,
and the switching locations translate to switching times between particular
modes of the hybrid system. Hence, this is similar to switching-time
optimization problems, e.g.,
\cite{egerstedt2006transition,shaikh2007hybrid,xu2004optimal} except that we
can only control a subset of mode switching times.

To describe an IPA treatment of the problem, we first present the hybrid
automaton model corresponding to the system operating on an optimal
trajectory.

\textbf{Hybrid automaton model}. We use a standard definition of a hybrid
automaton (e.g., see \cite{cassandras2007stochastic}) as the formalism to
model such a system. Thus, let $q\in Q$ (a countable set) denote the discrete
state (or mode) and $x\in X\subseteq\mathbb{R}^{n}$ denote the continuous
state. Let $\upsilon\in\Upsilon$ (a countable set) denote a discrete control
input and $u\in U\subseteq\mathbb{R}^{m}$ a continuous control input.
Similarly, let $\delta\in\Delta$ (a countable set) denote a discrete
disturbance input and $d\in D\subseteq\mathbb{R}^{p}$ a continuous disturbance
input. The state evolution is determined by means of $(i)$ a vector field
$f:Q\times X\times U\times D\rightarrow X$, $(ii)$ an invariant (or domain)
set $Inv:$ $Q\times\Upsilon\times\Delta\rightarrow2^{X}$, $(iii)$ a guard set
$Guard:$ $Q\times Q\times\Upsilon\times\Delta\rightarrow2^{X}$, and $(iv)$ a
reset function $r:$ $Q\times Q\times X\times\Upsilon\times\Delta\rightarrow
X$. The system remains at a discrete state $q$ as long as the continuous
(time-driven) state $x$ does not leave the set $Inv(q,\upsilon,\delta)$. If
$x$ reaches a set $Guard(q,q^{\prime},\upsilon,\delta)$ for some $q^{\prime
}\in Q$, a discrete transition can take place. If this transition does take
place, the state instantaneously resets to $(q^{\prime},x^{\prime})$ where
$x^{\prime}$ is determined by the reset map $r(q,q^{\prime},x,\upsilon
,\delta)$. Changes in $\upsilon$ and $\delta$ are discrete events that either
\emph{enable} a transition from $q$ to $q^{\prime}$ by making sure $x\in
Guard(q,q^{\prime},\upsilon,\delta)$ or \emph{force} a transition out of $q$
by making sure $x\notin Inv(q,\upsilon,\delta)$. We will classify all events
that cause discrete state transitions in a manner that suits the purposes of
IPA. Since our problem is set in a deterministic framework, $\delta$ and $d$
will not be used.%
\begin{figure*}
[ptb]
\begin{center}
\includegraphics[
scale=0.6
]%
{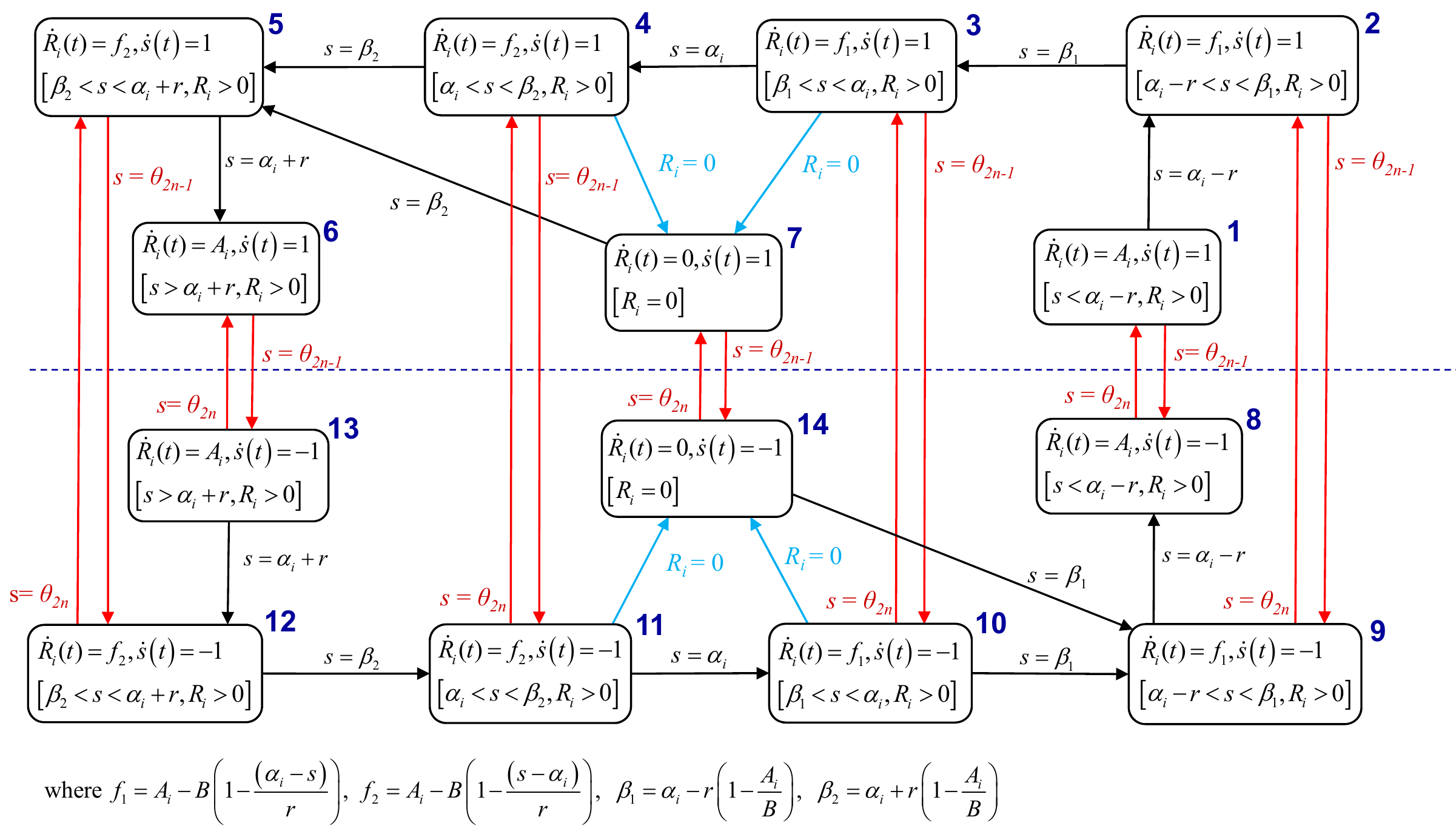}%
\caption{Hybrid automaton for each $\alpha_{i}$. Red arrows represent events
when the control switches between $1$ and $-1$. Blue arrows represent events
when $R_{i}$ becomes $0$. Black arrows represent all other events.}%
\label{fig:statetransition}%
\end{center}
\end{figure*}

We show in Fig. \ref{fig:statetransition} a partial hybrid automaton model of
the system: due to the size of the overall model, Fig.
\ref{fig:statetransition} is limited to the behavior of the agent with respect
to a single $\alpha_{i},i\in\{1,\ldots,M\}$. The model consists of 14 discrete
states (modes) and is symmetric in the sense that states $1-7$ correspond to
the agent operating with $u(t)=1$, and states $8-14$ correspond to the agent
operating with $u(t)=-1$. The events that cause state transitions can be
placed in three categories: $(i)$ The value of $R_{i}(t)$ becomes 0 and
triggers a switch in the dynamics of (\ref{eq:dynamicsR}). This can only
happen when $R_{i}(t)>0$ and $\dot{R}_{i}(t)=A_{i}-Bp_{i}(s(t))<0$ (e.g.,
in states $3$ and $4$), causing a transition to state $7$ in which the
invariant condition is $R_{i}(t)=0$. $(ii)$ The agent reaches a switching
location, indicated by the guard condition $s(t)=\theta_{j}$ for any
$j=1,\ldots,N$. In these cases, a transition results from a state $q$ to $q+7$
if $q=1,\ldots,6$ and to $q-7$ otherwise. $(iii)$ The agent position reaches
one of several critical values that affect the dynamics of $R_{i}(t)$ while
$R_{i}(t)>0$. When $s(t)=\alpha_{i}-r$, the value of $p_{i}(s(t))$ becomes
strictly positive and $\dot{R}_{i}(t)=A_{i}-Bp_{i}(s(t))>0$, as in the
transition $1\rightarrow2$. Subsequently, when $s(t)=\alpha_{i}-r(1-A_{i}%
/B)$, as in the transition $2\rightarrow3$, the value of $p_{i}(s(t))$
becomes sufficiently large to cause $\dot{R}_{i}(t)=A_{i}-Bp_{i}(s(t))<0$
so that a transition due to $R_{i}(t)=0$ becomes feasible at this state.
Similar transitions occur when $s(t)=\alpha_{i}$, $s(t)=\alpha_{i}%
+r(1-A_{i}/B)$, and $s(t)=\alpha_{i}+r$. The latter results in state $6$
where $\dot{R}_{i}(t)=A_{i}>0$ and the only feasible event is $s(t)=\theta
_{j}$, $j$ odd, when a switch must occur and a transition to state $13$ takes
place (similarly for state $8$).

\textbf{IPA review}. Before proceeding, we provide a brief review of the IPA
framework for general stochastic hybrid systems as presented in
\cite{cassandras2009perturbation}. In our case, the system is deterministic,
offering several simplifications. The purpose of IPA is to study the behavior
of a hybrid system state as a function of a parameter vector $\theta\in\Theta$
for a given compact, convex set $\Theta\subset\mathbb{R}^{l}$. Let $\{\tau
_{k}(\theta)\}$, $k=1,\ldots,K$, denote the occurrence times of all events in
the state trajectory. For convenience, we set $\tau_{0}=0$ and $\tau_{K+1}=T$.
Over an interval $[\tau_{k}(\theta),\tau_{k+1}(\theta))$, the system is at
some mode during which the time-driven state satisfies $\dot{x}\ =\ f_{k}%
(x,\theta,t)$. An event at $\tau_{k}$ is classified as $(i)$ \emph{Exogenous}
if it causes a discrete state transition independent of $\theta$ and satisfies
$\frac{d\tau_{k}}{d\theta}=0$; $(ii)$ \emph{Endogenous}, if there exists a
continuously differentiable function $g_{k}:\mathbb{R}^{n}\times
\Theta\rightarrow\mathbb{R}$ such that $\tau_{k}\ =\ \min\{t>\tau
_{k-1}\ :\ g_{k}\left(  x\left(  \theta,t\right)  ,\theta\right)  =0\}$; and
$(iii)$ \emph{Induced} if it is triggered by the occurrence of another event
at time $\tau_{m}\leq\tau_{k}$. Since the system considered in this paper does
not include induced events, we will limit ourselves to the first two event
types. IPA specifies how changes in $\theta$ influence the state $x(\theta,t)$
and the event times $\tau_{k}(\theta)$ and, ultimately, how they influence
interesting performance metrics which are generally expressed in terms of
these variables.

Given $\theta=[\theta_{1},\ldots,\theta_{N}]^{\mathtt{T}}$, we use the
notation for Jacobian matrices: $x^{\prime}(t)\equiv\frac{\partial
x(\theta,t)}{\partial\theta}$, $\tau_{k}^{\prime}\equiv\frac{\partial\tau
_{k}(\theta)}{\partial\theta}$, $k=1,\ldots,K$, for all state and event time
derivatives. It is shown in \cite{cassandras2009perturbation} that $x^{\prime
}(t)$ satisfies:
\begin{equation}
\frac{d}{dt}x^{\prime}\left(  t\right)  =\frac{\partial f_{k}\left(  t\right)
}{\partial x}x^{\prime}\left(  t\right)  +\frac{\partial f_{k}\left(
t\right)  }{\partial\theta} \label{xprimesolution}%
\end{equation}
for $t\in\lbrack\tau_{k},\tau_{k+1})$ with boundary condition:
\begin{equation}
x^{\prime}(\tau_{k}^{+})\ =\ x^{\prime}(\tau_{k}^{-})+\left[  f_{k-1}(\tau
_{k}^{-})-f_{k}(\tau_{k}^{+})\right]  \tau_{k}^{\prime}
\label{xprime boundary}%
\end{equation}
for $k=0,\ldots,K$. In addition, in (\ref{xprime boundary}), the gradient
vector for each $\tau_{k}$ is $\tau_{k}=0$ if the event at $\tau_{k}$ is
exogenous and
\begin{equation}
\tau_{k}^{\prime}=-\left[  \frac{\partial g_{k}}{\partial x}f_{k}(\tau_{k}%
^{-})\right]  ^{-1}\left(  \frac{\partial g_{k}}{\partial\theta}%
+\frac{\partial g_{k}}{\partial x}x^{\prime}(\tau_{k}^{-})\right)
\label{taukprime}%
\end{equation}
if the event at $\tau_{k}$ is endogenous and defined as long as $\frac
{\partial g_{k}}{\partial x}f_{k}(\tau_{k}^{-})\neq0$.

\textbf{IPA equations}. To clarify the presentation, we first note that
$i=1,\ldots,M$ is used to index the points where uncertainty is measured;
$j=1,\ldots,N$ indexes the components of the parameter vector; and
$k=1,\ldots,K$ indexes event times. In order to apply the three fundamental
IPA equations (\ref{xprimesolution})-(\ref{taukprime}) to our system, we use
the state vector $x\left(  t\right)  =[s\left(  t\right)  ,R_{1}%
(t),\ldots,R_{M}(t)]^{\mathtt{T}}$ and parameter vector $\theta=[\theta
_{1},\ldots,\theta_{N}]^{\mathtt{T}}$. We then identify all events that can
occur in Fig. \ref{fig:statetransition} and consider intervals $[\tau
_{k}(\theta),\tau_{k+1}(\theta))$ over which the system is in one of the 14
states shown for each $i=1,\ldots,M$. Applying (\ref{xprimesolution}) to
$s(t)$ with $f_{k}\left(  t\right)  =1$ or $-1$ due to (\ref{eq:agentdyn}) and
(\ref{u*}), the solution yields the gradient vector $\nabla s(t)=[\frac
{\partial s}{\partial\theta_{1}}(t),\ldots,\frac{\partial s}{\partial
\theta_{M}}(t)]^{\mathtt{T}}$, where
\begin{equation}
\frac{\partial s}{\partial\theta_{j}}(t)=\frac{\partial s}{\partial\theta_{j}%
}(\tau_{k}^{+}),\text{ for }t\in\lbrack\tau_{k},\tau_{k+1}) \label{sprime(t)}%
\end{equation}
for all $k=1,\ldots,K$, i.e., for all states $q(t)\in\{1,\ldots,14\}$.
Similarly, let $\nabla R_{i}(t)=[\frac{\partial R_{i}}{\partial\theta_{1}%
}(t),\ldots,\frac{\partial R_{i}}{\partial\theta_{M}}(t)]^{\mathtt{T}}$ for
$i=1,\ldots,M$. We note from (\ref{eq:dynamicsR}) that $f_{k}\left(  t\right)
=0$ for states $q(t)\in Q_{1}\equiv\{7,14\}$; $f_{k}\left(  t\right)  =A_{i}$
for states $q(t)\in Q_{2}\equiv\{1,6,8,13\}$; and $f_{k}\left(  t\right)
=A_{i}-Bp_{i}(s(t))$ for all other states which we further classify into
$Q_{3}\equiv\{2,3,11,12\}$ and $Q_{4}\equiv\{4,5,9,10\}$. Thus, solving
(\ref{xprimesolution}) and using (\ref{sprime(t)}) gives:
\begin{align*}
&  \nabla R_{i}\left(  t\right)  =\nabla R_{i}(\tau_{k}^{+})\\
&  -\left\{
\begin{array}
[c]{ll}%
0 & \text{if }q\left(  t\right)  \in Q_{1}\cup Q_{2}\\
B\left(  \frac{\partial p_{i}(s)}{\partial s}\right)  \nabla s\left(
\tau_{k}^{+}\right)  \cdot(t-\tau_{k}) & \text{otherwise}%
\end{array}
\right.
\end{align*}
where $\frac{\partial p_{i}(s)}{\partial s}=\pm\frac{1}{r}$ as evaluated from
(\ref{eq:linearmodel}) depending on the sign of $\alpha_{i}-s(t)$ at each
automaton state.

We now turn our attention to the determination of $\nabla s\left(  \tau
_{k}^{+}\right)  $ and $\nabla R_{i}(\tau_{k}^{+})$ from
(\ref{xprime boundary}), which involves the event time gradient vectors
$\nabla\tau_{k}=[\frac{\partial\tau_{k}}{\partial\theta_{1}},\ldots
,\frac{\partial\tau_{k}}{\partial\theta_{M}}]^{\mathtt{T}}$ for $k=1,\ldots
,K$.
Looking at Fig. \ref{fig:statetransition}, there are three readily
distinguishable cases regarding the events that cause state transitions:

\emph{Case 1}: An event at time $\tau_{k}$ which is neither $R_{i}=0$ nor
$s=\theta_{j}$, for any $j=1,\ldots,N$. In this case, it is easy to see that
the dynamics of both $s(t)$ and $R_{i}(t)$ are continuous, so that
$f_{k-1}(\tau_{k}^{-})=f_{k}(\tau_{k}^{+})$ in (\ref{xprime boundary}) applied
to $s\left(  t\right)  $ and $R_{i}(t),$ $i=1,\ldots,M$ and we get
\begin{equation}
\left\{
\begin{array}
[c]{l}%
\nabla s\left(  \tau_{k}^{+}\right)  =\nabla s\left(  \tau_{k}^{-}\right) \\
\nabla R_{i}(\tau_{k}^{+})=\nabla R_{i}(\tau_{k}^{-}),\text{ \ }i=1,\ldots,M
\end{array}
\right.
\end{equation}

\emph{Case 2}: An event $R_{i}=0$ at time $\tau_{k}$. This corresponds to
transitions $3\rightarrow7$, $4\rightarrow7$, $10\rightarrow14$ and
$11\rightarrow14$ in Fig. \ref{fig:statetransition} where the dynamics of
$s(t)$ are still continuous, but the dynamics of $R_{i}(t)$ switch from
$f_{k-1}(\tau_{k}^{-})=A_{i}-Bp_{i}(s(\tau_{k}^{-}))$ to $f_{k}(\tau
_{k}^{+})=0$. Thus, $\nabla s\left(  \tau_{k}^{-}\right)  =\nabla s\left(
\tau_{k}^{+}\right)  $, but we need to evaluate $\tau_{k}^{\prime}$ to
determine $\nabla R_{i}(\tau_{k}^{+})$. Observing that this event is
endogenous, (\ref{taukprime}) applies with $g_{k}=R_{i}=0$ and we get%
\[
\frac{\partial\tau_{k}}{\partial\theta_{j}}=-\frac{\frac{\partial R_{i}%
}{\partial\theta_{j}}\left(  \tau_{k}^{-}\right)  }{A_{i}-Bp_{i}%
(s(\tau_{k}^{-}))},\text{ }j=1,\ldots,N,\text{ }k=1,\ldots,K
\]
It follows from (\ref{xprime boundary}) that%
\[
\frac{\partial R_{i}}{\partial\theta_{j}}\left(  \tau_{k}^{+}\right)
=\frac{\partial R_{i}}{\partial\theta_{j}}\left(  \tau_{k}^{-}\right)
-\frac{[A_{i}-Bp_{i}(s(\tau_{k}^{-}))]\frac{\partial R_{i}}{\partial
\theta_{j}}\left(  \tau_{k}^{-}\right)  }{A_{i}-Bp_{i}(s(\tau_{k}^{-}%
))}=0
\]
Thus, $\frac{\partial R_{i}}{\partial\theta_{j}}\left(  \tau_{k}^{+}\right)  $
is always reset to $0$ regardless of $\frac{\partial R_{i}}{\partial\theta
_{j}}\left(  \tau_{k}^{-}\right)  $.

\emph{Case 3}: An event at time $\tau_{k}$ due to a control sign change at
$s=\theta_{j}$, $j=1,\ldots,N$. This corresponds to any transition between the
upper and lower part of the hybrid automaton in Fig. \ref{fig:statetransition}%
. In this case, the dynamics of $R_{i}(t)$ are continuous and we have
$\frac{\partial R_{i}}{\partial\theta_{j}}\left(  \tau_{k}^{+}\right)
=\frac{\partial R_{i}}{\partial\theta_{j}}\left(  \tau_{k}^{-}\right)  $ for
all $i,j,k$. On the other hand, we have $\dot{s}(\tau_{k}^{+})=u(\tau_{k}%
^{+})=-u(\tau_{k}^{-})=\pm1$. Observing that any such event is endogenous,
(\ref{taukprime}) applies with $g_{k}=s-\theta_{j}=0$ for some $j=1,\ldots,N$
and we get%
\begin{equation}
\frac{\partial\tau_{k}}{\partial\theta_{j}}=\frac{1-\frac{\partial s}%
{\partial\theta_{j}}\left(  \tau_{k}^{-}\right)  }{u(\tau_{k}^{-})}
\label{tauprime_n}%
\end{equation}
Combining (\ref{tauprime_n}) with (\ref{xprime boundary}) and recalling that
that $u(\tau_{k}^{+})=-u(\tau_{k}^{-})$, we have%
\[
\frac{\partial s}{\partial\theta_{j}}(\tau_{k}^{+})=\ \frac{\partial
s}{\partial\theta_{j}}(\tau_{k}^{-})+[u\left(  \tau_{k}^{-}\right)
-u(\tau_{k}^{+})]\frac{1-\frac{\partial s}{\partial\theta_{j}}\left(  \tau
_{k}^{-}\right)  }{u(\tau_{k}^{-})}=2
\]
where $\frac{\partial s}{\partial\theta_{j}}\left(  \tau_{k}^{-}\right)  =0$
because $\frac{\partial s}{\partial\theta_{j}}\left(  0\right)  =0=\frac
{\partial s}{\partial\theta_{j}}\left(  t\right)  $ for all $t\in\lbrack
0,\tau_{k})$, since the position of the agent cannot be affected by
$\theta_{j}$ prior to this event.

Now, let us consider the effect of perturbations to $\theta_{n}$ for $n<j$,
i.e., prior to the current event time $\tau_{k}$. In this case, we have
$\frac{\partial g_{k}}{\partial\theta_{n}}=0$ and (\ref{taukprime}) becomes%
\[
\frac{\partial\tau_{k}}{\partial\theta_{n}}=-\frac{\frac{\partial s}%
{\partial\theta_{n}}\left(  \tau_{k}^{-}\right)  }{u(\tau_{k}^{-})}%
\]
so that using this in (\ref{xprime boundary}) gives:%
\[
\frac{\partial s}{\partial\theta_{n}}(\tau_{k}^{+})=\frac{\partial s}%
{\partial\theta_{n}}(\tau_{k}^{-})-\frac{\left[  u\left(  \tau_{k}^{-}\right)
-u(\tau_{k}^{+})\right]  \frac{\partial s}{\partial\theta_{n}}\left(  \tau
_{k}^{-}\right)  }{u\left(  \tau_{k}^{-}\right)  }=-\frac{\partial s}%
{\partial\theta_{n}}\left(  \tau_{k}^{-}\right)
\]
Combining the above results, the components of $\nabla s(\tau_{k}^{+})$ where
$\tau_{k}$ is the event time when $s(\tau_{k})=\theta_{j}$ for some $j$, are
given by
\begin{equation}
\frac{\partial s}{\partial\theta_{n}}(\tau_{k}^{+})\ =\left\{
\begin{array}
[c]{ll}%
-\frac{\partial s}{\partial\theta_{n}}\left(  \tau_{k}^{-}\right)  & \text{if
}n=1,\ldots,j-1\\
2 & \text{if }n=j\\
0 & \text{if }n=j+1,\ldots,N
\end{array}
\right.  \label{sprimejump}%
\end{equation}
It follows from (\ref{sprime(t)}) and the analysis of all three cases above
that $\frac{\partial s}{\partial\theta_{j}}\left(  t\right)  $ for all $j$ is
constant throughout an optimal trajectory except at transitions caused by
control switching locations (\emph{Case 3}). In particular, for the $k$th
event corresponding to $s(\tau_{k})=\theta_{j}$, $t\in\lbrack\tau_{k},T]$, if
$u\left(  t\right)  =1$, then $\frac{\partial s}{\partial\theta_{j}}\left(
t\right)  =-2$ if $j$ is odd, and $\frac{\partial s}{\partial\theta_{j}%
}\left(  t\right)  =2$ if $j$ is even; similarly, if $u\left(  t\right)  =-1$,
then $\frac{\partial s}{\partial\theta_{j}}\left(  t\right)  =2$ if $j$ is odd
and $\frac{\partial s}{\partial\theta_{j}}\left(  t\right)  =-2$ if $j$ is
even. In summary, we can write $\frac{\partial s}{\partial\theta_{j}}\left(
t\right)  $ as
\begin{equation}
\frac{\partial s}{\partial\theta_{j}}\left(  t\right)  =\left\{
\begin{array}
[c]{cc}%
\left(  -1\right)  ^{j}\cdot2u\left(  t\right)  & t\geq\tau_{k}\\
0 & t<\tau_{k}%
\end{array}
\right.  \text{, \ }j=1,\ldots,N \label{sprime j}%
\end{equation}
Finally, we can combine (\ref{sprime j}) with our results for $\frac{\partial
R_{i}}{\partial\theta_{j}}\left(  t\right)  $ in all three cases above.
Letting $s(\tau_{l})=\theta_{j}$, we obtain the following expression for
$\frac{\partial R_{i}}{\partial\theta_{j}}\left(  t\right)  $ for all $k\geq
l,$ $t\in\lbrack\tau_{k},\tau_{k+1})$:
\begin{align}
&  \frac{\partial R_{i}}{\partial\theta_{j}}\left(  t\right)  =\frac{\partial
R_{i}}{\partial\theta_{j}}\left(  \tau_{k}^{+}\right)  \label{Rprime general}%
\\
+  &  \left\{
\begin{array}
[c]{cl}%
0 & \text{if }q(t)\in Q_{1}\cup Q_{2}\\
\left(  -1\right)  ^{j+1}\frac{2B}{r}u\left(  \tau_{k}^{+}\right)
\cdot(t-\tau_{k}) & \text{if }q(t)\in Q_{3}\\
-\left(  -1\right)  ^{j+1}\frac{2B}{r}u\left(  \tau_{k}^{+}\right)
\cdot(t-\tau_{k}) & \text{if }q(t)\in Q_{4}%
\end{array}
\right. \nonumber
\end{align}
with boundary condition
\begin{equation}
\frac{\partial R_{i}}{\partial\theta_{j}}(\tau_{k}^{+})=\left\{
\begin{array}
[c]{cl}%
0 & \text{if }q\left(  \tau_{k}^{+}\right)  \in Q_{1}\\
\frac{\partial R_{i}}{\partial\theta_{j}}(\tau_{k}^{-}) & \text{otherwise}%
\end{array}
\right.  \label{Rjump general}%
\end{equation}

\textbf{Objective Function Gradient Evaluation.} Since we are ultimately
interested in minimizing the objective function $J(\theta)$ (now a function of
$\theta$ instead of $u$) in (\ref{eq:costfunction}) with respect to $\theta$,
we first rewrite:%
\[
J(\theta)=\frac{1}{T}\sum_{i=1}^{M}\sum_{k=0}^{K}%
{\displaystyle\int_{\tau_{k}(\theta)}^{\tau_{k+1}(\theta)}}
R_{i}\left(  t,\theta\right)  dt
\]
where we have explicitly indicated the dependence on $\theta$. We then
obtain:
\begin{align*}
&  \nabla J(\theta)\\
&  =\frac{1}{T}%
{\displaystyle\sum_{i=1}^{M}}
{\displaystyle\sum_{k=0}^{N}}
\left(
{\displaystyle\int_{\tau_{k}}^{\tau_{k+1}}}
\nabla R_{i}\left(  t\right)  dt+R_{i}\left(  \tau_{k+1}\right)  \nabla
\tau_{k+1}-R_{i}\left(  \tau_{k}\right)  \nabla\tau_{k}\right)
\end{align*}
Observing the cancellation of all terms of the form $R_{i}\left(  \tau
_{k}\right)  \nabla\tau_{k}$ for all $k$, we finally get
\begin{equation}
\nabla J(\theta)=\frac{1}{T}%
{\displaystyle\sum_{i=1}^{M}}
{\displaystyle\sum_{k=0}^{N}}
{\displaystyle\int_{\tau_{k}}^{\tau_{k+1}}}
\nabla R_{i}\left(  t\right)  dt. \label{dJdtheta}%
\end{equation}
The evaluation of $\nabla J(\theta)$ therefore depends entirely on $\nabla
R_{i}\left(  t\right)  $, which is obtained from (\ref{Rprime general}%
)-(\ref{Rjump general}) and the event times $\tau_{k}$, $k=1,\ldots,K$, given
initial conditions $s\left(  0\right)  =0$, $R_{i}\left(  0\right)  $ for
$i=1,\ldots,M$ and $\nabla R_{i}(0)=0$.

\textbf{Objective Function Optimization.} We now seek to obtain $\theta
^{\star}$ minimizing $J(\theta)$ through a standard gradient-based
optimization scheme of the form%

\begin{equation}
\theta^{l+1}=\theta^{l}-\eta_{l}\tilde{\nabla}J(\theta^{l})
\label{eq:updatetheta}%
\end{equation}
where $\{\eta_{l}\}$ is an appropriate step size sequence and $\tilde{\nabla
}J(\theta)$ is the projection of the gradient $\nabla J(\theta)$ onto the
feasible set (the set of $\theta$ satisfying the constraint
\eqref{eq:thetaconstraint}). The optimization scheme terminates when
$|\tilde{\nabla}J(\theta)|<\varepsilon$ (for a fixed threshold $\varepsilon$)
for some $\theta$. Our IPA-based algorithm to obtain $\theta^{\star}$
minimizing $J(\theta)$ is summarized in Alg. \ref{alg:IPA} where we have
adopted the Armijo step-size (see \cite{polak1997optimization}) for
$\{\eta_{l}\}$.
\begin{algorithm}
\caption{: IPA-based optimization algorithm to find $\theta^{\star}$}
\label{alg:IPA}
\begin{algorithmic}[1]
\STATE Set $N=\lfloor\frac{T}{L}\rfloor$ ($\lfloor\cdot \rfloor$ is the floor function), and set $\theta=[\theta_{1},\ldots,\theta_{N}]^{\tt T}$ satisfying constraint \eqref{eq:thetaconstraint}
\REPEAT
\STATE Compute $s(t)$,  $t\in[0,T]$ using $\theta$
\STATE Compute $\tilde\nabla J(\theta)$ and update $\theta$ through \eqref{eq:updatetheta}
\UNTIL{$|\tilde\nabla J(\theta)|<\epsilon$}
\IF{$\theta$ satisfies Prop. \ref{lem:switchingpoints}}
\STATE {\bf Stop}, return $\theta$ as $\theta^{\star}$
\ELSE
\STATE Set $N+1\rightarrow N$ and set $\theta_{N}=s(T)$
\STATE Go to Step $2$
\ENDIF
\end{algorithmic}
\end{algorithm}

Recalling that the dimension $N$ of $\theta^{\star}$ is unknown (it depends on
$T$), a distinctive feature of Alg. \ref{alg:IPA} is that we vary $N$ by
possibly increasing it after a vector $\theta$ locally minimizing $J$ is
obtained, if it does not satisfy the necessary optimality condition in Prop.
\ref{lem:switchingpoints}. We start
the search for a feasible $N$ by setting it to $\lfloor\frac{T}{L}\rfloor$,
the minimal $N$ for which $\theta$ can satisfy Prop. \ref{lem:switchingpoints}%
, and only need to increase $N$ if the locally optimal $\theta$ vector violates Prop.
\ref{lem:switchingpoints}.

It is possible to increase $N$ further after Alg. \ref{alg:IPA} stops, and
obtain a local optimal $\theta$ vector with a lower cost. This is due to
possible non-convexity of the problem in terms of $\theta$ and $N$. In
practice, this computation can take place in the background while the agent is
in operation. Alternatively, we can adapt a receding horizon formulation to
compute the optimal control on-line. This approach is explained in more detail
in Sec. \ref{sec:extensions}.

\section{Numerical results\label{sec:exper}}

\begin{figure}[ptb]
\centering
\includegraphics[scale=.4]%
{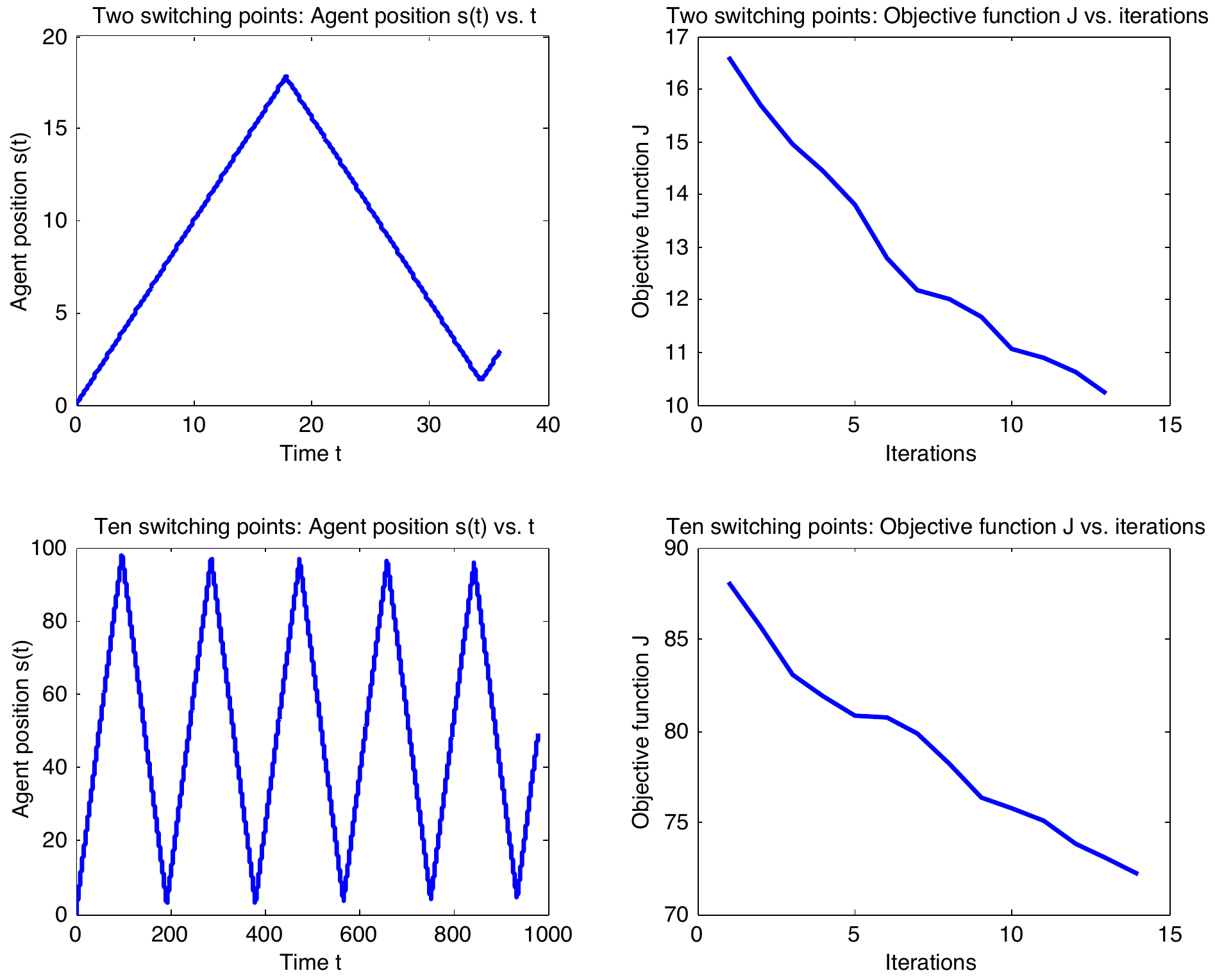}%
\caption{Numerical results. Top figures correspond to $L=20$, $T=36$, $21$ sampling points in $[0,L]$.  Bottom figures correspond to $L=100$, $T=980$, $101$ sampling points in $[0,L]$.  Left plots: optimal trajectories. Right plots: $J$ versus iterations. }
\label{fig:results}
\end{figure}

In this section we present two numerical examples where we have used Alg.
\ref{alg:IPA} to obtain an optimal persistent monitoring trajectory. The
results are shown in Fig. \ref{fig:results}. The top two figures correspond to
an example with $L=20$, $M=21$, $\alpha_{1}=0$, $\alpha_{M}=20$, and the
remaining sampling points are evenly spaced between each other. Moreover,
$A_{i}=0.01$ for all $i$, $B=3, r=4, R_{i}(0)=2$ for all $i$ and $T=36$. We start the
algorithm with $\theta=[12]^{\tt T}$ and $\varepsilon=2\times10^{-10}$. The
algorithm stopped after 13 iterations (about $9$ sec) using Armijo step-sizes,
and the cost, $J$, was decreased from $16.63$ to $J^{\star}=10.24$ with
$\theta^{\star}=[17.81,1.29]^{\tt T}$, i.e., the dimension increased by $1$. In
the top-left, the optimal trajectory $s^{\star}(t)$ is plotted; in the
top-right, $J$ is plotted against iterations. We also increased $N$ to $3$
with initial $\theta=[12,16,4]$; Alg. \ref{alg:IPA} converged to a local
minimum $J=13.27>J^{\star}=10.24$ under $N=2$.

The bottom two figures correspond to an example with $L=100$, $M=101$ and
evenly spaced sampling points over $[0,L]$, $A_{i}=0.01$ for all $i$, $B=3,$ $r=4,$
$R_{i}(0)=2$ for all $i$ and $T=980$. We start the algorithm with $N=9$,
$\theta=[95,95,95,95,95,5,5,5,5]^{\tt T}$ and same $\varepsilon$. The algorithm
stopped after 14 iterations (about $10$ min, an indication of the rapid
increase in computational complexity) using Armijo step-sizes, and $J$ was
decreased from $88.10$ to $J^{\star}=70.49$ with  $\theta^{\star}%
=[98.03,96.97,96.65,96.35,95.70,2.94,3.21,3.61,4.08,4.57]^{\tt T}$ where $N=10$.
Note that the cost is much higher in this case due to the larger number of
sampling points. Moreover, none of the optimal switching locations is at $0$
or $L$, consistent with Prop. \ref{lem:switchingpoints}. We also increased $N$
to $11$ with $\theta=[90,90,90,90,90,90,10,10,10,10,10]$; Alg. \ref{alg:IPA}
converged to $101.56>J^{\star}=70.49$ under $N=10$.

\section{Extensions}
\label{sec:extensions}
In this section we briefly discuss extensions to a ``myopic'' Receding Horizon (RH) framework, or a setting with multiple agents.  Our proposed uncertainty model can be directly used to solve the persistent monitoring problem with a RH approach by solving \textbf{Problem P1} not for the time horizon $T$, but for a smaller time window $H$, where $H\leq T$, repeatedly every time interval $h\leq H$.  Because $H$ is usually much smaller than $T$, and since the optimal control is shown to be ``bang-bang'' when not inside a singular arc, it can be assumed that the control is constant (denoted as $u$) during the horizon $[t, t+H]$.  In this case, the problem of minimizing the cost function \eqref{eq:costfunction} over $u\in [-1,1]$ is a scalar optimization problem and its solution can be obtained explicitly, given the initial conditions of $s(t)$ and $R_{i}(t)$.  The RH controller operates as follows: at time $t$, the optimal control is computed for $[t, t+H]$ and is used for the time interval $[t,t+h]$.  This process is repeated every $h$ units of time, until $t=T$.  In our numerical examples, the cost obtained using the RH framework is very close to the optimal cost (consistently within $5\%$), and since an explicit solution is available, the optimal control can be computed quickly and in real-time. The RH framework can also accommodate situations where events are triggered in real-time at some sampling points; in the virtual queue analogy, this means the inflow rates $A_{i}$ of some queues are time-varying.

This approach also opens up future work for multiple agents in 2-D or 3-D mission spaces.  In a multi-agent framework, we can use the same model for uncertainty, but with a joint event detection probability function $p(x,s_{1},\ldots,s_{n})$, where there are $n$ agents.  This joint probability can be expressed in terms of individual detection probabilities $p(x,s_{i})$ as: $p(x,s_{1},\ldots,s_{n})=1-\prod_{i=1}^{n}(1-p(x,s_{i}))$.  Although the optimal control problem can still be fully solved for multiple agents in the 1-D mission space, this problem quickly becomes intractable in higher dimensions.   In this case, we aim to develop a unified receding horizon approach that integrates with our previous cooperative coverage control strategies \cite{li2006cooperative}.

\section{Conclusions}
\label{sec:concl}
We have formulated a persistent monitoring problem where we consider a dynamic environment with uncertainties at points changing depending on the proximity of the agent.   We obtained an optimal control solution that minimizes the accumulated uncertainty over the environment, in the case of a single agent and 1-D mission space.  The solution is characterized by a sequence of switching points, and we use an IPA-based gradient algorithm to compute the solution.   We also discussed extensions of our approach using a receding horizon framework.   Ongoing work aims at solving the problem with multiple agents and a richer dynamical model for each agent, as well as addressing the persistent monitoring problem in 2-D and 3-D mission spaces.

\bibliographystyle{plain}
\bibliography{./Papers}

\end{document}